\documentclass[final,twocolumn,5p,12pt]{elsarticle}
\usepackage[T1]{fontenc}
\usepackage{graphicx}	
\usepackage{amsmath,amssymb}
\usepackage{times}
\usepackage{units}
\usepackage[english]{babel}
\usepackage[utf8]{inputenc}
\usepackage{booktabs,array}
\usepackage{listings}
\usepackage{xspace}
\usepackage{fp}
\usepackage{ifthen}
\usepackage{color}

\def \be {\begin{equation}}
\def \ee {\end{equation}}
\def \bear {\begin{eqnarray}}
\def \eear {\end{eqnarray}}

\journal{Computational Material Science}
\begin{document}
\begin{frontmatter}

\title{Growth of two-dimensional Au patches in graphene pores: \\a density-functional study}

\author{Saku Antikainen}
\author{Pekka Koskinen\corref{cor}}
\ead{pekka.koskinen@iki.fi}
\address{NanoScience Center, Department of Physics, University 
of Jyvaskyla, 40014 Jyvaskyla, Finland}

\cortext[cor]{Corresponding author}

\begin{abstract}
Inspired by recent studies of various two-dimensional (2D) metals such as Au, Fe and Ag, we study the growth of two-dimensional gold patches in graphene pores by density-functional theory. We find that at room temperature gold atoms diffuse readily on top of both graphene and two-dimensional gold with energy barriers less than $0.5$ eV. Furthermore, gold atoms move without barriers from the top of graphene to its edge and from the top of 2D gold to its edge. The energy barriers are absent even at the interface of 2D gold and graphene, so that the gold atoms move effortlessly across the interface.  We hope our demonstration for the propensity of diffusing gold atoms to grow 2D gold patches in graphene pores will inspire the fabrication of these patches experimentally.
\end{abstract}

\end{frontmatter}


\section{Introduction}
The great success with graphene has sparked much additional interest to other possible two-dimensional (2D) materials. The dimensionality can change the properties of the material greatly, as evidenced particularly well by graphene: it has extremely high carrier mobility and thermal conductivity, and it demonstrates the Quantum Hall effect.\cite{yang2, zhang} Another example is the transition-metal dichalcogenide \(\text{MoS}_2\): bulk \(\text{MoS}_2\) is an indirect bandgap semiconductor, while a monolayer \(\text{MoS}_2\) is a direct gap semiconductor.\cite{mak, splendiani} Both graphite and bulk \(\text{MoS}_2\) consist of covalently bound layers that are held together by the weak van der Waals (vdW) interactions. Bulk metals have no such layered structures, and thus the fabrication of two-dimensional metallic structures is more problematic. However, the recent interest in 2D metals has triggered several studies to inspect the possibility of their existence.

For example, the simulations of 2D metals have shown promising results about their stability. For gold Yang et al. have predicted a stable, two-dimensional lattice structure with hexagonal symmetry.\cite{yang} In addition, bond strength was found to increase greatly when going from bulk 3D Au to 2D Au, analogously to the case of 3D diamond and 2D graphene. Similar predictions were made for 2D silver, which also was found to prefer a hexagonal lattice structure.\cite{yang2} In addition to static properties, Koskinen and Korhonen have predicted the existence of a liquid phase in a free-standing, atomically thin 2D Au layer suspended by graphene pores.\cite{pekka} However, a free-standing two-dimensional layer of gold is yet to be produced experimentally. Zhao et al. have managed to create a single-atom-thick iron layer suspended in graphene pores\cite{zhao}; this idea could be likewise applied to other metals. Shao \emph{et al.} suggest with their simulations that a square lattice monolayer of Fe is energetically unstable, and that the experimentally observed Fe monolayers would instead be made of a mixture of Fe and C.\cite{shao} The possibility of a combination of carbon and metal is certainly worth investigating in further studies of 2D metals.

In any case, earlier studies indicate that gold would be a particularly suitable candidate for a 2D metal. In nanoscale, gold has been found to behave very differently from the inert bulk gold. For example, small gold clusters of sizes up to 20 atoms have been shown to exhibit catalytic activity in the combustion of CO\cite{sanchez}, and gold cluster anions of sizes as large as 11 atoms have been shown to have two-dimensional ground states.\cite{johansson} This exceptional planar stability has been attributed to the relativistic effects on gold.\cite{hakkinen} The studies of 2D Au have been promising, but creating a free-standing 2D monolayer of gold is experimentally problematic. Yet it has been shown that a gold atom interacts strongly with a graphene edge.\cite{wang} These properties of gold, combined with the success of producing an Fe monolayer suspended in graphene pores, give faith that two-dimensional structures could be synthesized with gold and perhaps even with other metals.

Here we have used density-functional theory (DFT) to investigate how Au atoms behave at the graphene edge and thus to model the growth of a 2D gold patch in a graphene pore. We performed DFT calculations to investigate the behavior of gold atoms in four different stages of the growth process: i) originating from some source of atomic gold, gold atoms move on top of a graphene sheet, ii) a gold patch begins to form at the edge of the graphene, iii) gold atoms move on top of the gold patch, and iv) gold atoms move from the top of the gold patch to its edge, thereby growing the patch. We studied the adsorption of gold at different adsorption sites on top of the graphene and 2D gold sheets, as well as at the edge of the sheets. In addition, we determined the potential energy surfaces (PES) for the movement of gold atoms between various adsorption sites. These potential energy landscapes indicate clearly that gold atoms prefer to move quickly from the top of graphene across the graphene-gold interface and to the edge of the 2D gold patch.

\section{Methods}

\begin{figure*}[!tb]
  \centering
  \includegraphics[width=1.0\textwidth]{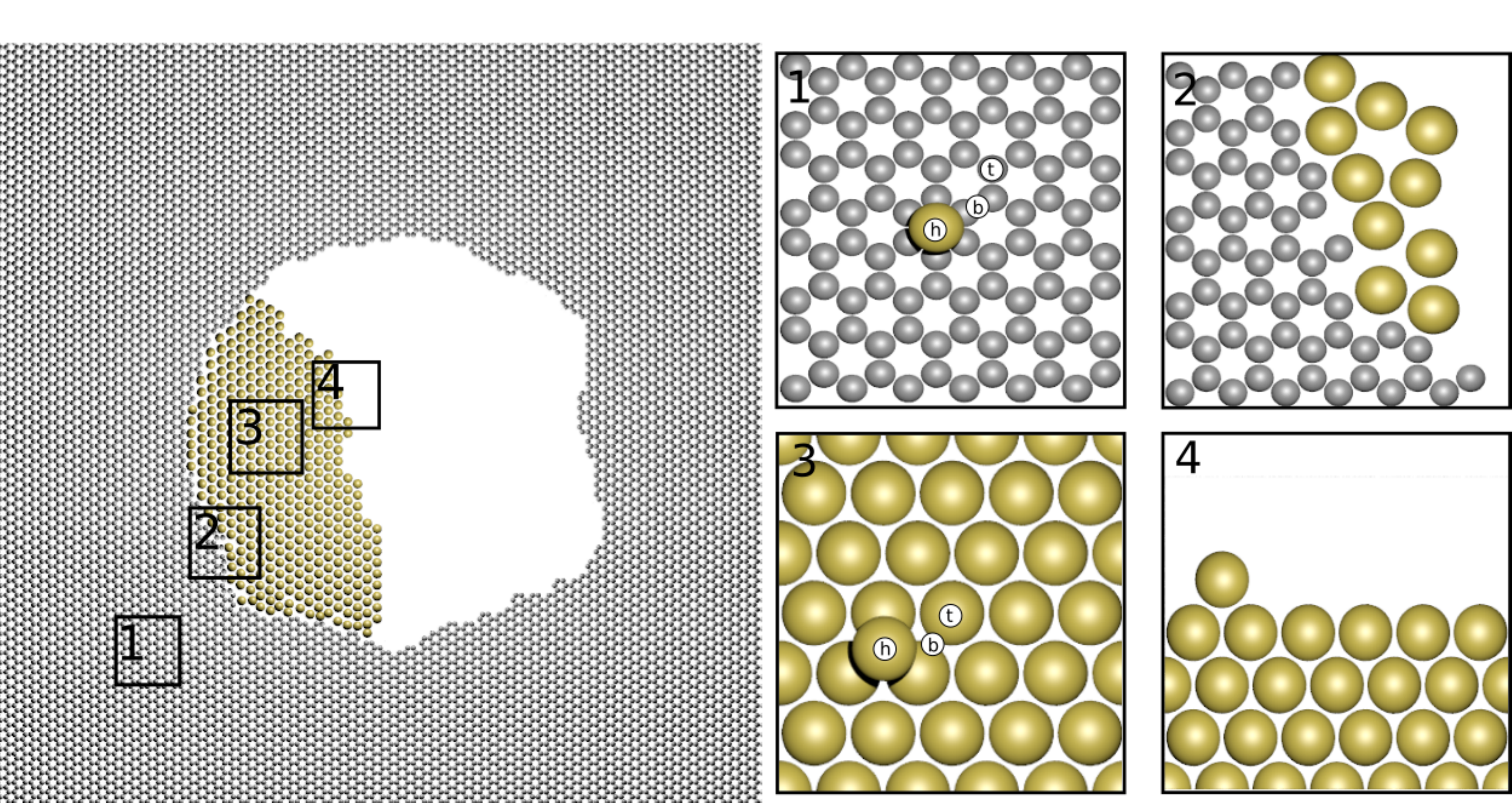}
  \caption{Overall picture of the growth process. On the left is an infinite 2D graphene sheet with a growing patch of gold in the middle. The growth process is divided into four pieces, labeled 1-4: 1) a 2D periodic graphene sheet with top[t], hollow[h] and bridge[b] adsorption sites, 2) a 1D periodic graphene zigzag-edge and armchair-edge with gold at the edge, 3) a 2D periodic gold patch with top[t], hollow[h] and bridge[b] adsorption sites, and 4) a 1D periodic gold edge.}
  \label{fig:overall}
\end{figure*}

Our goal was to model the growth of a gold patch at the edge of a graphene pore. More precisely, we modeled gold atoms moving on top of graphene towards the edge so that when they met the edge a gold patch began to form. The overall process was broken into four stages, sketched in Figure \ref{fig:overall}. First, we used a 2D graphene sheet with a single gold atom at various adsorption sites. Second, to model the growth of graphene-2D metal interface we used graphene nanoribbon with varying number of gold atoms at the edge. We used both a zig-zag-edged graphene nanoribbon (ZGNR) and an armchair-edged graphene nanoribbon (AGNR). Third, we used a 2D gold sheet with a single gold atom at various adsorption sites. And fourth, to model the actual growth of the gold patch, we used a one-dimensional gold edge with an additional gold atom at the edge.

The simulations were performed in the atomic simulation environment \cite{ase} and using the density-functional code GPAW \cite{gpaw1,gpaw2}, which is based on the projector-augmented wave method (PAW) \cite{paw}. The generalized gradient approximation exchange-correlation functional of Perdew, Burke, and Ernzerhof (PBE)\cite{pbe} was used throughout the calculations. All calculations were made in local basis mode (LCAO) with double-zeta polarized basis, and some additional calculations in Finite Difference (FD) mode.

For the 2D structures (graphene and 2D Au), the convergence of adsorption energy of a single Au atom on the 2D sheet was tested with respect to unit cell size and k-point sampling. The adsorption energy was calculated according to the equation
\begin{equation} \label{eq:ads}
  E_{\text{ads}} = E_{\text{2D }} + E_{\text{Au atom}} - E_{\text{relax}},
\end{equation}
where \(E_{\text{2D }}\) is the energy of the 2D sheet without the adsorbate, \(E_{\text{Au atom}}\) is the energy of a free Au atom and \(E_{\text{relax}}\) is the energy of the relaxed system. To get \(E_{\text{relax}}\), the atoms of the 2D sheet were fixed, while the adsorbate was allowed to move until the forces on all atoms were \textless 0.05 eV/\(\text{\AA}\). The calculations were made with both LCAO- and FD-mode, but we chose LCAO-mode for the rest of the simulations because its accuracy turned out to be sufficient compared to FD-mode. We chose k-point sampling of $3\times 3\times 1$ for both of the systems; for 2D Au we chose cell size of $4\times 4$ atoms and for graphene $8\times 4$ atoms, as the adsorption energy was found to be sufficiently converged already at these values. We used a lattice constant of 1.42 \(\text{\AA}\) for graphene and 2.76 \(\text{\AA}\) for 2D Au; the 2D Au had a hexagonal lattice structure.\cite{yang} All the structures were non-periodic in z-direction with a 6.0 \(\text{\AA}\) vacuum on both sides.

The edges of gold and graphene were modeled using 1D periodic systems. The periodicity was in x-direction with 6.0 \(\text{\AA}\) of vacuum in both y- and z-directions. For the 1D structures (1D Au, ZGNR, and AGNR), the convergence of adsorption energy of an Au atom at the edge with respect to the number of rows of Au  or C-atoms in the y-direction was confirmed. In these convergence calculations the atom positions were kept fixed because at this stage we were merely interested in the convergence of the adsorption energy with respect to the electronic structure rather than the relaxation of the atoms. The edge energy of the nanoribbon was obtained from the equation for the total energy of the nanoribbon:
\begin{equation} \label{eq:E_w}
  E_{\text{total}} = -N \cdot \varepsilon_{\text{2D }} + L_{\text{edge}} \cdot \varepsilon_{\text{edge}},
\end{equation}
where \(N\) is the number of atoms in the unit cell, \(\varepsilon_{\text{2D }}\) is the cohesion energy of the infinite 2D sheet, \(L_{\text{edge}}\) is the total edge length (2 times the cell x-length) and \(\varepsilon_{\text{edge}}\) is the edge energy. We chose the number of atoms in a row to be 12 for AGNR and 10 for ZGNR, as this produced nearly equal unit cell sizes and allowed for enough gold atoms to be placed at the edge. Finally, to keep the cell sizes comparable, the unit cell of 1D Au contained four gold atoms in a row.

To study the actual growth of the gold patch, we added Au atoms one by one to the edge and allowed the system to relax between each added atom using the Broyden-Fletcher-Goldfarb-Shanno (BFGS) algorithm for optimization\cite{broyden}. For all the systems described above, we calculated the adsorption energy of a gold atom at different sites. In addition, we studied the potential energy surface (PES) of gold atom on the 2D sheets and along the edges. To calculate the potential energy surfaces between various adsorption sites, we used the nudged elastic band -method (NEB)\cite{neb} with 3 images between the start and end points, which was sufficient because the reaction paths were fairly simple.

\section{Results}
\subsection{Stage 1: Gold on graphene}
We begun to model the growth process by investigating the movement of a single gold atom on top of graphene. This is a good starting point because much is already known of the diffusion of gold on graphene.\cite{amft2,malola} The obtained adsorption energies of a single Au atom at hollow, bridge, and top sites on 2D graphene were calculated from Eq.\eqref{eq:ads}. The adsorption energies are very low, and top-site has a slightly higher energy (102 meV) than the bridge-site (96 meV) or the hollow-site (73 meV). The study of Amft \emph{et al.}\cite{amft} report similar numbers (top-site 99 meV and bridge-site 81 meV) with the exception of hollow site, where no binding was predicted. Along with PBE, they also tested other functionals to account for the vdW interaction between the Au atom and graphene. While the introduction of vdW forces increased the adsorption energies, the order of the energies remained the same, and the top site remained energetically most favorable. With or without accounting for vdW forces, their calculations showed that the likely diffusion path is along the C-C bonds. Although vdW interactions are frequently important in 2D materials, the usage of PBE is justified because here the main point of interest is the growth process of gold; within the scope of our work the effect of a vdW functional would be minor.

\begin{figure}[!b]
  \centering
  \includegraphics[width=1.0\linewidth]{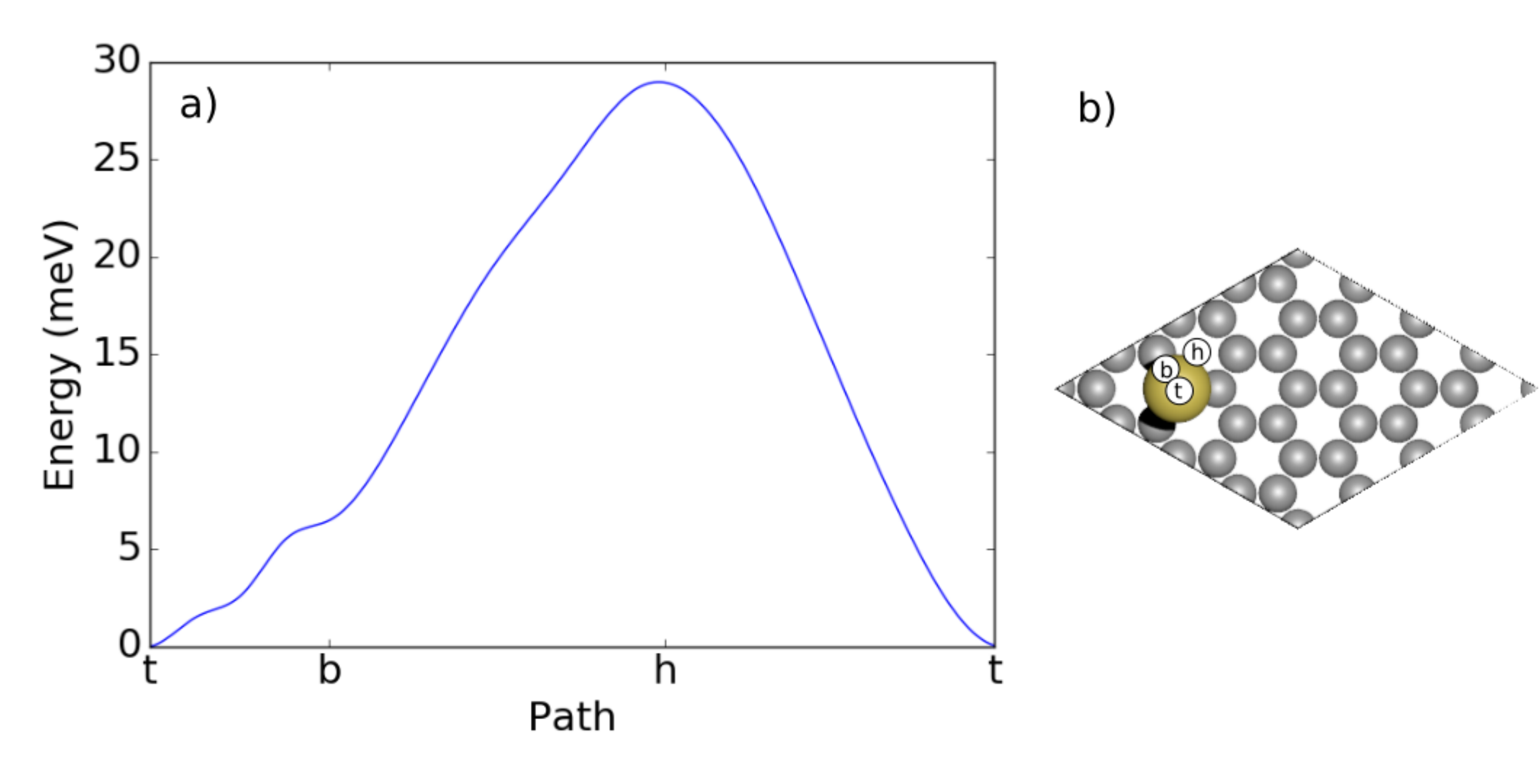}
  \caption{a) Potential energy surface (PES) of Au atom on top of graphene and b) the unit cell used in calculations, with hollow [h], bridge [b], and top [t] sites.}
  \label{fig:neb_gra}
\end{figure}

We studied the diffusion of Au atom on graphene for three different paths: top-bridge (t-b), bridge-hollow (b-h) and hollow-top (h-t), and the results can be seen in Fig. \ref{fig:neb_gra}. No energy barriers were found on any of these separate paths. As the difference of energies between top and bridge sites is very low (6 meV), it is reasonable to expect a gold atom to move readily on top of graphene from top site to top site along the bridges; the same conclusion was reached in the aforementioned study of Amft \emph{et al.}\cite{amft}

\subsection{Stage 2: Gold at graphene edge}
\begin{figure}[!tb]
  \centering
  \includegraphics[width=1.0\linewidth]{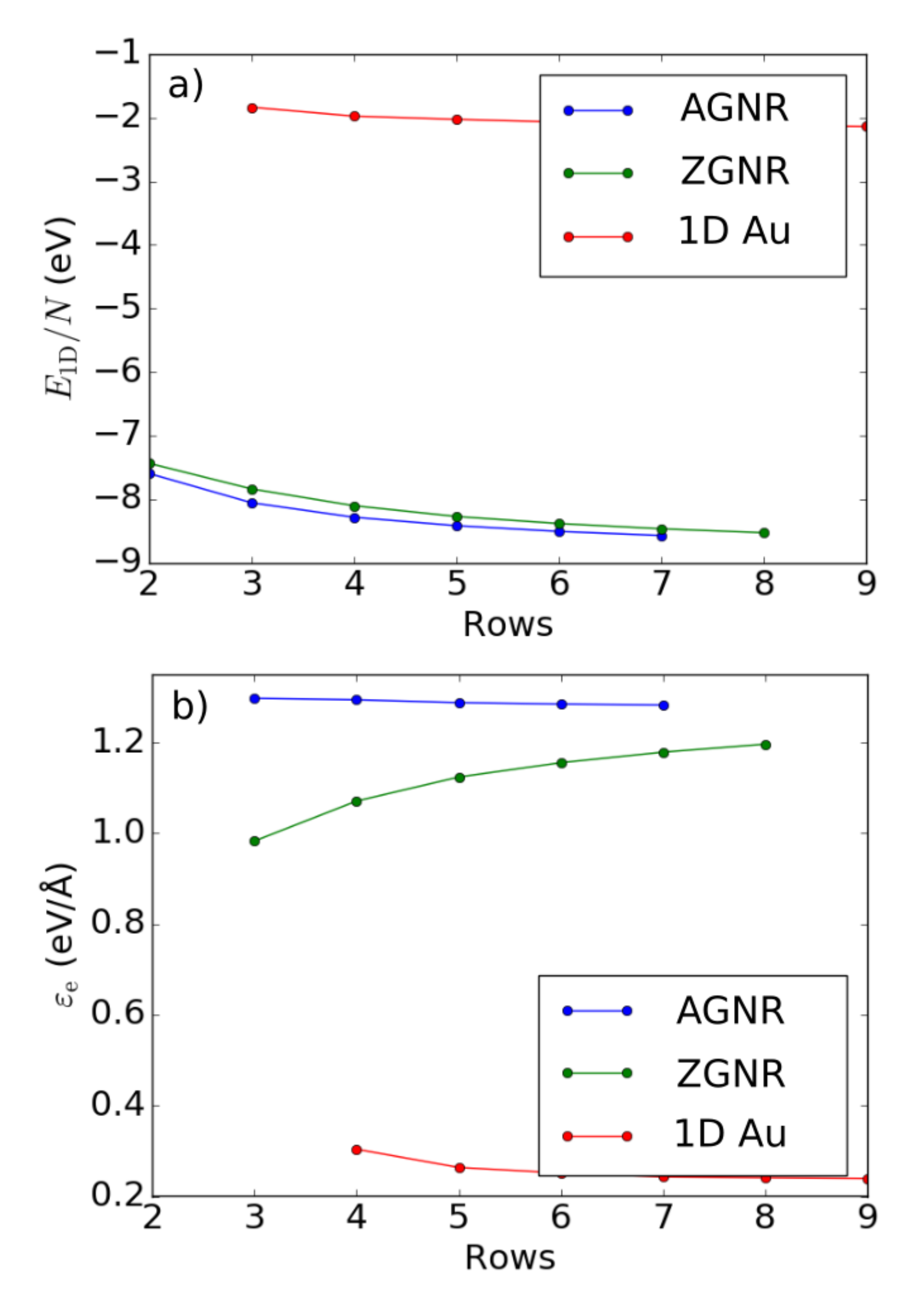}
  \caption{a) Energy per atom of 1D gold, AGNR and ZGNR as a function of atomic rows. b) Edge energy of 1D gold, AGNR and ZGNR as a function of atomic rows.}
  \label{fig:w_E}
\end{figure}

To investigate the growth of gold patch at graphene edge we constructed zigzag and armchair graphene nanoribbons with 2-7 rows of C-atoms. Each row contained in AGNRs 12 C atoms and in ZGNRs 10 C atoms. In Fig. \ref{fig:w_E}a the energy per atom is calculated and plotted as a function of rows. We fitted the curve of Eq. \eqref{eq:E_w} for 3-8 rows and thus obtained the edge energies, which can be seen in Fig. \ref{fig:w_E}b. In their simulations of graphene nanoribbons, Koskinen \emph{et al.} found the edge energies \(\varepsilon_{\text{edge}}^{\text{ac}} = 0.98 \text{ eV/\AA}\) and \(\varepsilon_{\text{edge}}^{\text{zz}} = 1.31 \text{ eV/\AA}\).\cite{pekka2} Our present results are in fair agreement with the earlier numbers, considering that we did no optimization at this point and that we fixed the atom positions using constant bond length 1.42 \(\text{\AA}\) of 2D graphene.

Next we added a single gold atom at two different adsorption sites near the edge: at a hollow site on the edge (in plane) and at a top site on top of the edge (out of plane). Three rows of C-atoms were used with fixed bottom-row atoms. The systems were allowed to relax and the adsorption energies were calculated. As a result, the top sites turned out to be unstable, as during the optimization the Au atoms moved spontaneously to a hollow site at the edge. From the edge adsorption energies (\(E_{\text{ads}}^{\text{ac}} = 5.62\) eV and \(E_{\text{ads}}^{\text{zz}} = 4.61\) eV) we see that the binding is much stronger at the edge than on top of the graphene, which is expected due to the available dangling bonds of edge carbon atoms. This is in good agreement with previous literature.\cite{wang}

\begin{figure}[!tb]
  \centering
  \includegraphics[width=1.0\linewidth]{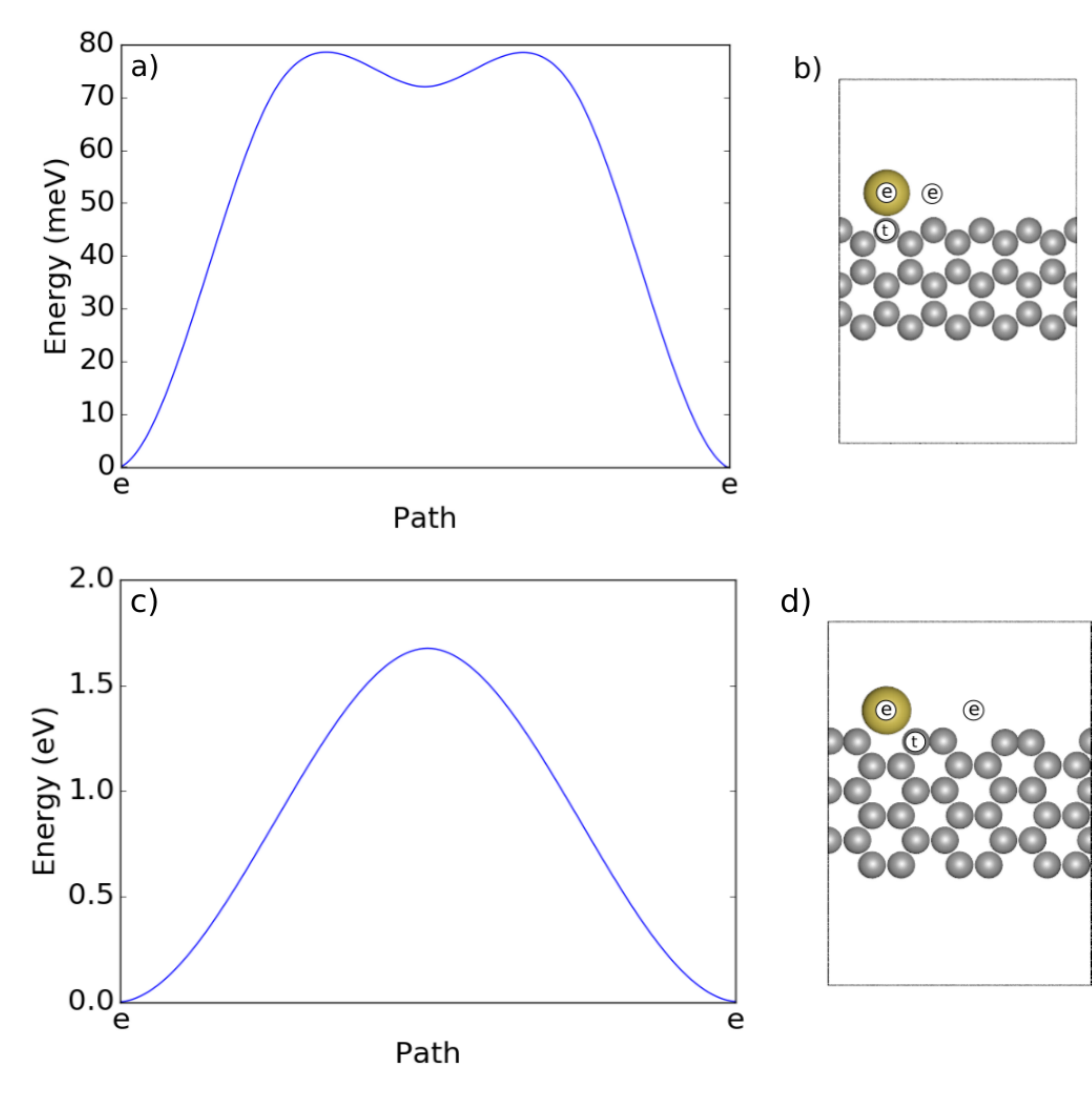}
  \caption{a) PES of Au atom moving along ZGNR edge and b) the unit cell used in ZGNR calculations, with top [t] and edge [e] sites. c) PES of Au atom moving along AGNR edge and d) the unit cell used in AGNR calculations, with top [t] and edge [e] sites.}
  \label{fig:neb_GNR}
\end{figure}

Next we studied the movement of the Au atom along the AGNR and ZGNR edges. The results of energy calculations on a path between two edge sites are shown in Fig.~\ref{fig:neb_GNR}. AGNR shows much higher energy barriers (\(\sim\)1.7 eV) than ZGNR (\(\sim\)80 meV). This might be attributed to the two carbon atoms that the Au atom will have to move across on the AGNR path as opposed to one on the ZGNR path. In other words, in zigzag graphene nanoribbons the dangling bonds are equidistant, whereas in armchair graphene nanoribbons they are separated alternatingly by longer and shorter distances.

\begin{figure*}[!tb]
  \centering
  \includegraphics[width=0.95\textwidth]{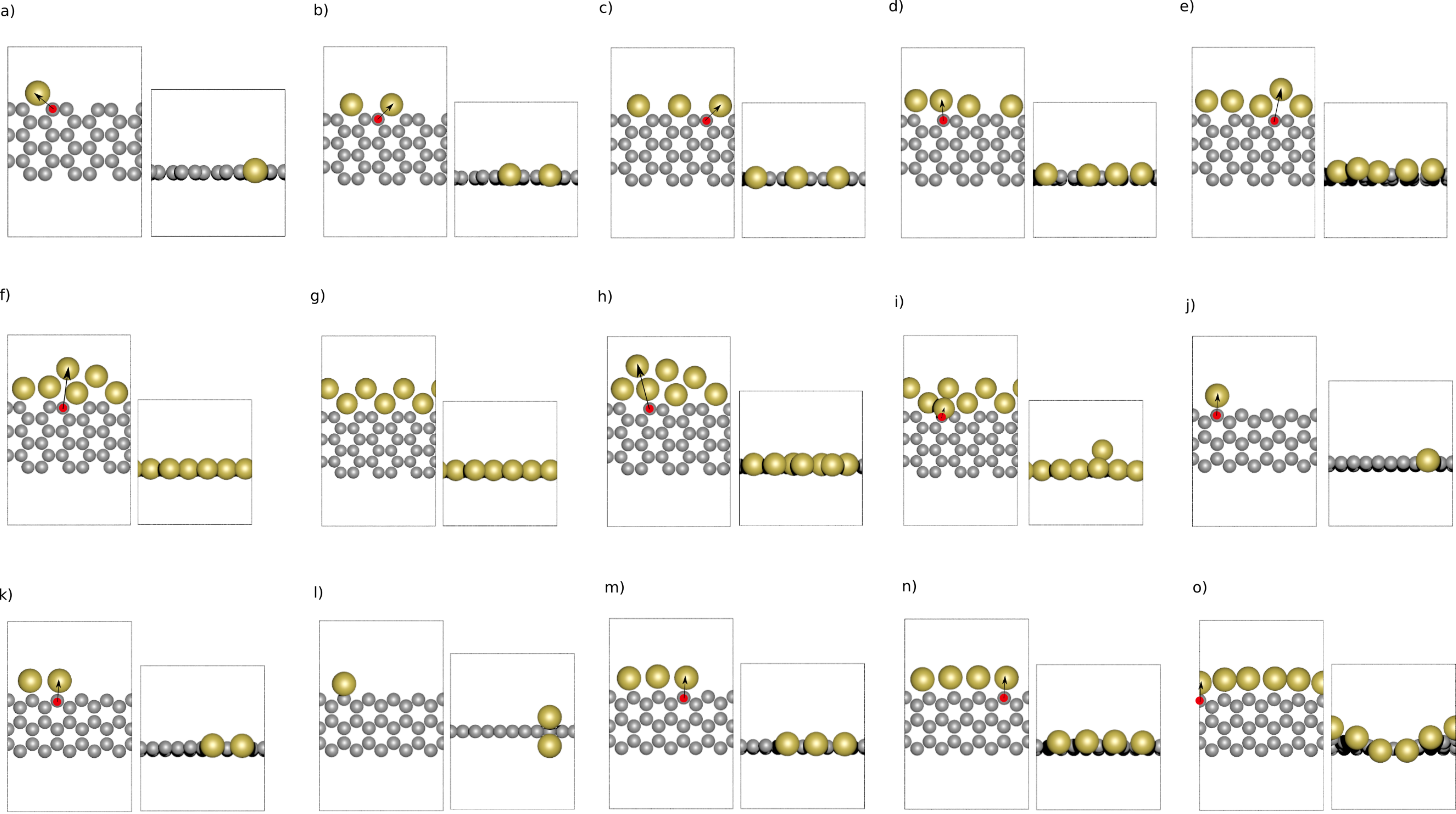}
  \caption{Unit cells of AGNR and ZGNR systems with various amounts of gold at the edge, with top and side views for each system. The red circle indicates the position in which the most recent Au atom was added (the atom that models the actual growth), and the arrow points to the final position after optimization.}
  \label{fig:geo_edge_full}
\end{figure*}

Next we begun to model stepwise the formation of the 2D gold patch to the graphene edge by adding gold atoms one by one. The gold atoms were added to various top sites near the graphene edge (an example is seen in Fig. \ref{fig:neb_GNR}b). A total of 7 atoms were added to AGNR edge and 5 atoms to ZGNR; all optimized systems are shown in Fig. \ref{fig:geo_edge_full}. In AGNR, the first three gold atoms settled for the edge sites (such as ones shown in Fig. \ref{fig:neb_GNR}d). The fourth atom also fit in the same row, but the fifth and sixth atoms started forming a second row of 2D gold. The seventh atom replaced another Au atom in the first row and nudged it to the second row (Fig. \ref{fig:geo_edge_full}h). In other words, the 2D gold patch did not necessarily grow from the edge, but it could grow also at the graphene-gold interface by nudging previously settled gold atoms farther away. 

For comparison, we also studied different 6- and 7-Au atom systems at armchair edge (Fig. \ref{fig:geo_edge_full}g and \ref{fig:geo_edge_full}i, respectively). Here the six atoms were placed in different starting positions before optimization. This six-atom system (Fig. \ref{fig:geo_edge_full}g) was found to have lower energy than the one-atom-at-a-time grown system of Fig. \ref{fig:geo_edge_full}f. Again with the addition of the 7th Au atom, we found no energy barrier when moving to the edge. Interestingly, while the unit cells were kept roughly the same size with both AGNR and ZGNR, all 5 atoms fit in the first row with ZGNR. However, this came at the expense of large out-of-plane distortions (Fig. \ref{fig:geo_edge_full}o). For ZGNR, we also studied an additional tetragonal geometry, where two gold atoms were bound out of the GNR plane at the edge (Fig. \ref{fig:geo_edge_full}l). This system was found to have higher energy than the other two-atom ZGNR system (Fig. \ref{fig:geo_edge_full}k) by 2.65 eV. Thus, energetically the interface prefers planar growth.

\subsection{Stage 3: Gold atom on 2D gold}
After investigating the graphene-2D gold interface, we moved on to investigate a gold atom moving on top of 2D gold sheet. The adsorpion energies were obtained for an Au atom on top of hollow, bridge and top sites. Here we found that the hollow site has the highest adsorption energy (1.37 eV), bridge site is quite close (1.28 eV) and top site has the lowest adsorption energy (0.91 eV). Compared to the case of gold on graphene, adsorption energies are much higher, which can be understood by the different nature of chemical bonding. And while on graphene Au preferred the top site, here it preferred the hollow site.

\begin{figure}[!tb]
  \centering
  \includegraphics[width=0.99\linewidth]{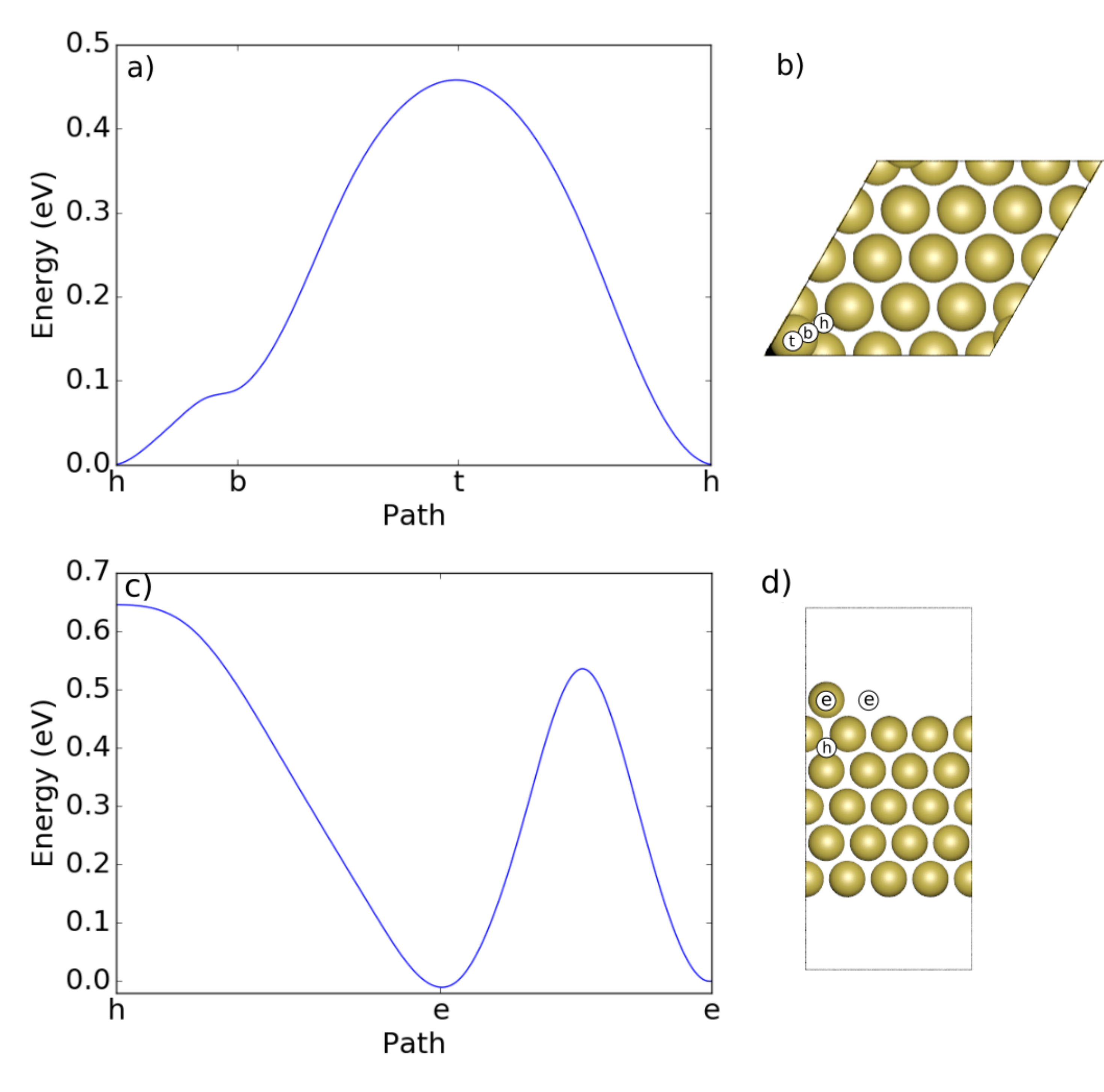}
  \caption{a) PES of Au atom on top of 2D gold and b) the unit cell used in 2D gold calculations, with top [t], bridge [b] and hollow [h] sites. c) PES of Au atom near the 1D gold edge and d) the unit cell used in 1D gold calculations, with hollow [h] and edge [e] sites.}
  \label{fig:neb_Au}
\end{figure}

To illustrate the movement of the Au atom on 2D Au, we once again performed potential energy surface calculations, the results of which can be seen in Fig. \ref{fig:neb_Au}a. Here the Au atom follows the path hollow-bridge-top-hollow. PES suggests that Au atom moving on top of 2D Au would most likely hop between the hollow sites via bridge sites, with a mere 90 meV energy barrier. Compared to graphene, the energy barriers here are still higher, especially on the paths involving the top site.

\subsection{Stage 4: Gold at the edge of 2D gold}
As the final stage, we investigated a gold atom at the edge of 2D gold patch that was modeled by a 1D gold ribbon. As with graphene, a 1D Au edge was constructed with 3-9 atomic rows, and the energy per atom was calculated and plotted with respect to the number of rows (Fig. \ref{fig:w_E}a). The resulting edge energies are shown in Fig. \ref{fig:w_E}b.

We studied the adsorption of a single Au atom on the edge, with two adsorption sites of interest: hollow site on top of the 1D Au, and an edge site on the side. The calculations were performed with 5 rows of Au atoms while the two bottom rows were fixed during the optimization. As a result, the adsorption energy at the edge site (2.27 eV) was found to be considerably higher than at the hollow site (1.62 eV). It is notable that the optimization of an Au atom at a hollow site close to the edge brought the Au atom much closer to the edge, almost to a bridge site.

The movement of the Au atom at the edge was studied along two paths, along a path from the hollow site on top of the 1D Au to the edge site (h-e) and along a path between two edge sites (e-e). The results of the calculations are shown in Fig. \ref{fig:neb_Au}c. The atom was found to hop from the top to the edge easily; the energy barrier was practically absent. Yet the movement along the edge however came with an energy barrier of \(\sim\)0.55 eV.

\section{Discussion}

\begin{table*}[!t]
\centering
\caption{Au atom adsorption energies \(E_{\text{ads}}\) (eV) and distances \(d\) (\(\text{\AA}\)) to nearest atoms at different adsorption sites on graphene, 2D Au, AGNR, ZGNR and 1D Au. }
\label{tab:ads}
\begin{tabular}{lllllllll}\toprule
 & \multicolumn{2}{c}{hollow} & \multicolumn{2}{c}{bridge} & \multicolumn{2}{c}{top} & \multicolumn{2}{c}{edge} \\
  & \(E_{\text{ads}}\) & \(d\)    & \(E_{\text{ads}}\) & \(d\) & \(E_{\text{ads}}\) & \(d\) & \(E_{\text{ads}}\) & \(d\) \\
  \hline
1 Graphene   & 0.073 & 3.57      & 0.096  & 3.19  & 0.102 & 3.08   & - & -\\
2 AGNR & - & - & - & - & - & - & 5.62 & 2.13 \\
2 ZGNR & - & - & - & - & - & - & 4.61 & 1.97 \\
3 2D Au    & 1.37 &  2.81      & 1.28 & 2.75    & 0.91  & 2.67 & - & - \\
4 1D Au & 1.62 & 2.71 & - & - & - & - & 2.27 & 2.67 \\
\hline
\end{tabular}
\end{table*}

\begin{figure}[!tb]
  \centering
  \includegraphics[width=0.8\linewidth]{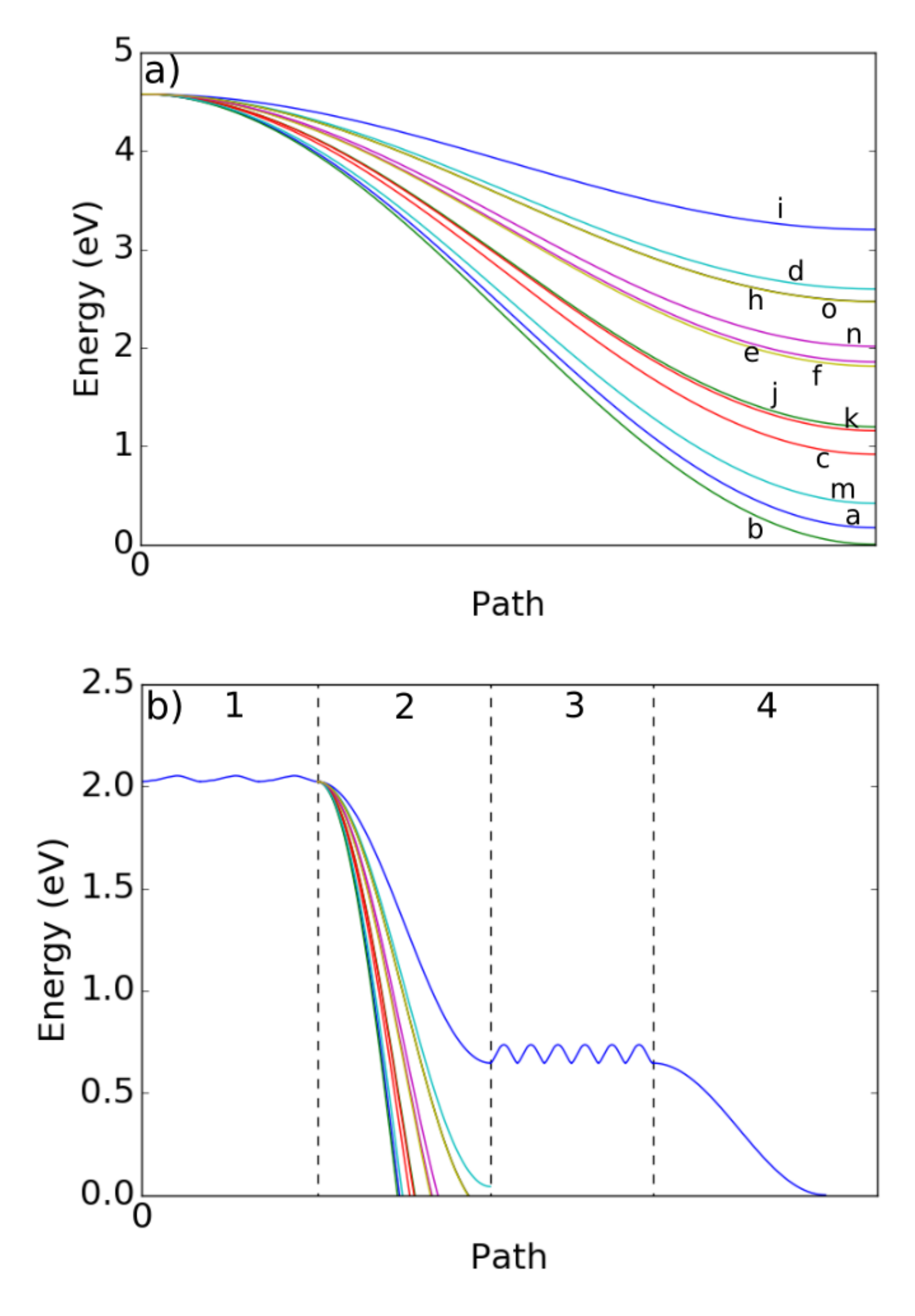}
  \caption{a) PES of Au atom moving from the top of graphene to the edge with different number of gold atoms at the edge. Each graph represents a different system as seen in Fig. \ref{fig:geo_edge_full} (the energy on top of graphene is set equal for different systems, according to system in Fig.~\ref{fig:geo_edge_full}b). b) A sketch for the overall PES for an Au atom contributing to the growth of the golden patch. Numbers 1-4 indicate the four stages shown in Fig. \ref{fig:overall}, with PES of stage 2 detailed in panel a. The blue path thus illustrates one possible PES for a gold atom during the growth process.}
  \label{fig:path_full3}
\end{figure}

The adsorption energies and the distances to the nearest neighboring atom of a single gold atom on various adsorption sites are summarized in Table \ref{tab:ads}. On top of graphene gold shows weak adsorption, while at the edge of graphene gold shows strong adsorption. There is also another possible adsorption site between the edge sites of ZGNR, as seen in Fig. \ref{fig:neb_GNR}. But because the energy barrier separating the sites is low (\textless 10 meV), the lower-energy edge sites are much more likely to get occupied by the incoming gold atoms.

The overall picture of the growth process is quite clear, at least when viewed via the potential energy surface. Figure \ref{fig:path_full3}b illustrates one possible PES for a gold atom during its traversing to the edge. The extremely low energy barrier (\(\sim\)30 meV) shows that a gold atom is likely to move readily on top of graphene. Moreover, it continues to move from the top to the edge without an energy barrier. The same trend continues even when more gold atoms are added. Interestingly, there is also a possibility for the new gold atoms to replace old gold atoms in the first row and nudge them farther to the second row, implying that patches can grow also at the graphene-gold interface.

Furthermore, as with graphene, the energy barrier of a gold atom moving on top of 2D gold is very low (90 meV on hollow-bridge-hollow path), and no energy barrier was found when moving to the edge of 2D gold. While atoms move to the edge effortlessly, there still exists energy barriers when moving along the edge. The armchair edge has particularly large barrier when moving along the edge (1.7 eV). 
As the gold patch grows, empty space closes up, until eventually the entire pore becomes filled. Our simulations indicate that diffusion will be easy up to the last atom. The resulting patch geometry will be either atomically flat or slightly non-planar, depending on the room available for the final atom\cite{pekka}. Ideally the growth would stop once the patch fully fills the pore, so that additional Au atoms would avoid getting trapped on top of the patch and thus avoid growing more layers on top of the 2D Au. 

Such accurate growth control requires low concentration of Au atoms. Low concentration has been an implicit assumption throughout the paper, because it has justified the investigations of the growth one atom at a time. High concentrations would mean that the diffusing atoms would sinter, form dimers and larger clusters, which would slow down the diffusion and hamper the growth. Computational investigations of the diffusion processes or edge growth would be prohibitively complex due to the sheer number of possible diffusion and reaction pathways and are therefore beyond the scope of this paper.\cite{freund_2008} Most important, larger clusters would be more prone to trigger the growth of 3D instead of 2D patches. Thus, the experimental synthesis of stable 2D patches is most likely to succeed at the low concentration limit.

In addition to high concentration, high temperature is another factor that could cause the collapse of the 2D gold patch into a 3D nanoparticle. This happens because of entropic factors, regardless of the favoring of 2D structures by pure energetics. Zhao \emph{et al.} observed this kind of collapse for a 2D Fe patch when the Fe particles were under prolonged electron irradiation; they found that the Fe membrane-armchair graphene interface remained stable the longest, compared to Fe -zigzag graphene interface.\cite{zhao} Nevertheless, the 2D Fe membranes remained stable for several minutes under the irradiation. The low energy barriers of our study (\textless 0.5 eV) indicate that room temperature (300 K) should be enough for rapid diffusion and patch growth. It can be expected that the effects of temperature might be the greatest during the stage 2 of our study, at the gold-graphene interface, where the potential energy drops are the largest.

We note that the choice of exchange-correlation functional affects more adsorption energies and less diffusion energy barriers. Since our main interest here is on energy barriers, our choice to exlude vdW interactions should be a reasonable approximation, as demonstrated in the study of Amft \emph{et al.}\cite{amft} Overall, the low energy barriers and the ease of the growth of the gold patch parhaps could be anticipated, but our study does further clarify the picture of the interaction of gold and graphene edge.

In summary, we have studied the growth of 2D gold at graphene pores, modeled by graphene zigzag and armchair edges. Gold was chosen because of its known 2D stability. Yet, earlier studies have demonstrated that also other metals show fast diffusion on graphene and strong binding to graphene defects\cite{gan_2008,tang_2012,yang_2016}. Our follow-up articles will concentrate on these other metals, as well as other pore materials, such as hexagonal boron nitride, transition metal dichalcogenides, and black phosphorus.\cite{geim_2013} Other pore materials are worth investigating, although it is an evident risk that the pores with more comples edge morphologies pose serious challenges for the stable growth of atomically thin 2D patches.\cite{pastewka_2013} But although we focused only on Au and graphene pores, we hope this study clarified the microscopic processes during the growth of a 2D metal patch in graphene pores and encourages experiments to push the limits for the patch sizes of stable 2D metal patches. As shown by this study, gold makes an excellent candidate for this because of its low diffusion barriers and strong binding with the graphene edge; the chances for experimental realization at room temperature should be fair.

\section{Acknowledgements}
We acknowledge the Academy of Finland for funding and CSC - IT Center for Science in Finland for computer resources.

\end{document}